# The CERN Detector Safety System for the LHC Experiments

S. Lüders
*CERN EP/SFT, 1211 Geneva 23, Switzerland; Stefan.Lueders@cern.ch*

R.B. Flockhart, G. Morpurgo, S.M. Schmeling
*CERN IT/CO, 1211 Geneva 23, Switzerland*

The Detector Safety System (DSS), currently being developed at CERN under the auspices of the Joint Controls Project (JCOP), will be responsible for assuring the protection of equipment for the four LHC experiments. Thus, the DSS will require a high degree of both availability and reliability. After evaluation of various possible solutions, a prototype is being built based on a redundant Siemens PLC front-end, to which the safety-critical part of the DSS task is delegated. This is then supervised by a PVSS SCADA system via an OPC server. The PLC front-end is capable of running autonomously and of automatically taking predefined protective actions whenever required. The supervisory layer provides the operator with a status display and with limited online reconfiguration capabilities. Configuration of the code running in the PLCs will be completely data driven via the contents of a "Configuration Database". Thus, the DSS can easily adapt to the different and constantly evolving requirements of the LHC experiments during their construction, commissioning and exploitation phases.

## 1. INTRODUCTION

The Detector Safety System (DSS) project covers one of the grey areas that still existed in the development process of the LHC[1] experiments at CERN: equipment protection. For the last large CERN accelerator project (LEP[2]) experiment safety on all levels was ensured by one single system, whereas for the LHC it was decided to delegate the responsibility for the highest level of safety[3] to a CERN-wide system, the CERN Safety System (CSS). Normal operation of the detectors is performed using the corresponding detector control systems (DCS), which also handles low level safety, e.g. the overheating of crates. This left an area of uncertainty, especially as the availability and reliability of a PC-based control system does not seem to be sufficient to ensure proper equipment protection.

At the beginning of 2001, the four LHC experiments produced a working document defining requirements for a system assuring equipment protection for the valuable, and sometimes irreplaceable, detectors. The outcome of this is the DSS.

## 2. SCOPE AND GOALS OF A DETECTOR SAFETY SYSTEM

The main goal of the DSS is to detect abnormal and potentially harmful situations, and to minimize the consequent damage to the experimental equipment by taking "protective actions". By implementing this strategy, a reduction of the occurrence of higher level alarms with more serious consequences can be expected, and therefore an increase of the experiment's running time and efficiency. The Detector Safety System should complement and not duplicate existing systems, such as the Detector Control System and the CERN Safety System. By working together in a complementary manner, these three systems will ensure that situations that may lead to equipment damage, or place people in danger, are well covered.

## 3. REQUIREMENTS AND GENERAL ARCHITECTURE

As a consequence of the above mentioned goals, the following main requirements were defined for the DSS. It has to be

- highly reliable and available, as well as simple and robust,
- provide a cost-effective solution for experimental safety,
- operate permanently and independently of the state of DCS and CSS,
- able to take immediate actions to protect the equipment,
- scalable, so that it may evolve with the experiments during their assembly, commissioning, operation and dismantling (a time-span of approximately 20 years),
- maintainable over the lifetime of the experiments,
- configurable, so that changes in the setup can be accounted for,
- able to connect to all sub-systems and sub-detector safety systems, and
- integrated into the DCS, so that existing tools can be reused, the same look & feel is presented to all users, and the monitoring, logging and presenting are standardized.

Taking all these requirements into account, the DSS will consist of three main entities:

---
[1] *L*arge *H*adron *Co*llider, accelerator currently being built at CERN.

[2] The *L*arge *E*lectron *P*ositron collider finished physics running in 2000.

[3] According to CERN rules there are three alarm levels (1-3); level 3 is defined as "accident or serious abnormal situation, especially where people's lives are, or may be, in danger" [1].

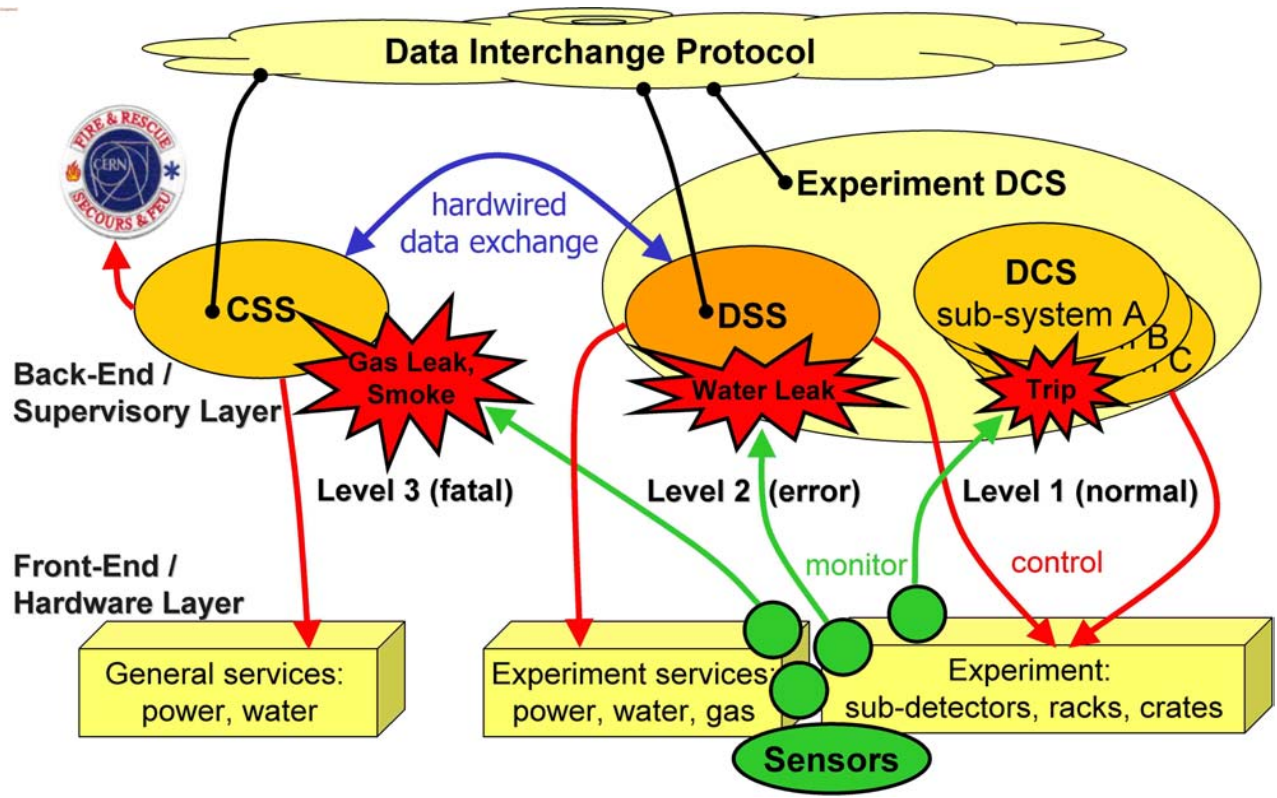

Figure 1: How the Detector Safety System integrates within the experimental control infrastructure.

1. The "Back-end", which constitutes the User Interface of the DSS, and which is based on the PVSS[4] SCADA[5] system and the JCOP Framework tools, is used for configuration, monitoring and logging, user interfaces, and as gateway to external information.
2. The "Front-end", to which the safety critical part of the DSS task is delegated, is built of highly reliable hardware, and can act upon hardwired information coming from dedicated sensors or from external systems (e.g. CSS). It is composed of several Detector Safety Units (DSU). Each DSU is responsible for a distinct geographical area. The inter-communication between the Front-end and the Back-end is performed through a dedicated OPC server / gateway.
3. The DSS sensors, which collect temperature, humidity and other physical parameters, as well as status information from various sub-detectors and external systems.

A complete overview of the DSS inside the detector control system architecture is shown in
Figure 1. It is composed of the following entities:

- the equipment that is acted upon by DCS, DSS, and CSS, subdivided here into two parts, the primary services (e.g. power, water, and primary gas supply) and the equipment under control of the experiment (e.g. sub-detectors, racks, and gas systems),
- the DCS, which is a coherent control system running PVSS on the supervisory and sub-system levels together with their own front-end parts. This includes the infrastructure control system, which controls racks, etc., together with its own sensors, e.g. temperature sensors inside the racks. The DCS might take corrective action to maintain normal operation.
- the CSS together with its own sensors, taking all required safety actions (e.g. calling the fire brigade) in case of an alarm of level 3. The CSS is required by law, and conforms to relevant International, European, and National standards.
- the DSS, consisting of a back-end and a front-end together with its own sensors and which is embedded in the Experiment's DCS,
- the Technical Services, providing the primary services (power, water, and gas),
- the Data Interchange Protocol (DIP), that provides information exchange between the experiments, the LHC machine, the technical services, and the CSS.

---

[4] *Prozeß-Visualisierungs- und Steuerungs-System* made by ETM AG, Austria; www.etm-ag.at
[5] *Supervisory Control And Data Acquisition*

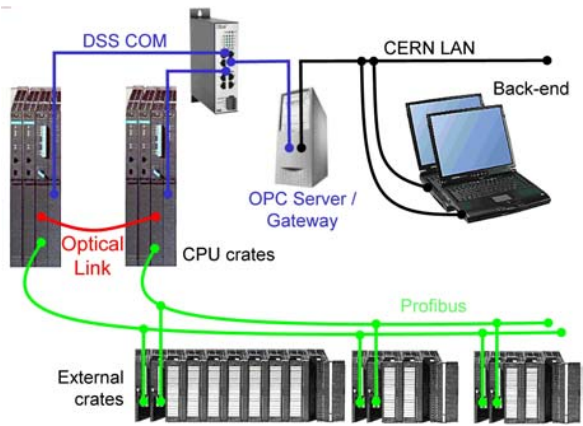

Figure 2: General DSS Hardware Layout

## 4. IMPLEMENTATION

The DSS implementation distinguishes between a Front-end, to which the safety-critical part of the DSS task is delegated, and a Back-end, which supervises the Front-end though a SCADA system. Their inter-communication is performed through a dedicated OPC[6] server / gateway. The general layout for a typical experiment is shown Figure 2.

### 4.1. Front-end Architecture

The Front-end must be capable of running autonomously and of automatically taking predefined protective actions whenever required. Its design must be as simple and robust as possible, but at the same time highly reliable and available.

The standard used for such applications in industry is "Programmable Logic Controllers" (PLCs). The DSS implementation follows this approach, using a redundant PLC system from Siemens[7] which is certified for Safety Integrity Level (SIL) 2 applications. In such systems, the two CPUs constantly monitor and compare their internal states and their communication lines, and automatically detect abnormalities. In the event of a problem, only the "good" branch continues to operate.

The status of the experiment's equipment is continually monitored by digital and analog sensors (e.g. PT100 for temperature measurement, humidity sensors, and status signals of sub-detectors). All values are filtered and checked. Required actions are determined automatically and immediately by the PLC and taken by DSS actuators (in this case, usually by cutting the electrical power). The sensors and actuators are dedicated to the DSS. Only hardwired sensors are considered to be safe, since networked information can not be guaranteed to be reliable.

All parts of the Front-end are compliant to the relevant safety norms.

#### 4.1.1. The PLC

The core of the DSS Front-end is the redundant PLC system S7-400 H from Siemens[8]. Each CPU of the redundancy pair scans the input signals of the DSS digital and analog inputs attached to the PLC system. Both CPUs run the same DSS Process Code and determine the state of the outputs (the DSS actuators) according to a set of rules defined by the user – the "Alarm-Action-Matrix". This sequence (scanning, processing, and output control) is repeated periodically and is named the "PLC cycle". The cycle time for a typical experiment is of the order of 500 ms, depending on the complexity of the Alarm-Action-Matrix and the number of sensors and actuators. Thus the reaction time of the DSS to an event is below one second.

Parts of the DSS Process Code and the corresponding data blocks can be modified without disturbing the DSS operation itself. Therefore, the DSS software can evolve to cover future needs.

The CPU internal state, the input values and the results of intermediate processing steps of both CPUs are monitored and compared regularly during a PLC cycle, by the redundancy process. Differences result in a shutdown of the faulty CPU, which does not affect the DSS itself as the second CPU continues to perform the DSS monitoring task.

Each of the two redundant CPUs is mounted in a crate together with redundant power supplies and Ethernet communication. To achieve synchronization, Siemens interconnects the two CPUs via four (non-redundant) optical fibers. Common time synchronization of both CPUs is achieved by connecting them to the network time server at CERN via NTP[9].

Several external crates make up the interface to the DSS sensors and actuators through I/O modules. Each external crate can hold up to eight I/O modules. Since both, sensors and actuators are widely spread over caverns and the surface of the experiment's site, the external crates act as cable concentrators to minimize sensor/actuator cable length.

Up to 32 external crates can be connected to both CPUs via a redundant cable pair using the PROFIbus protocol. Thus, redundancy is achieved down to the level of PROFIbus communication. The I/O modules themselves are not redundant, but sensors or actuators can be connected in a redundant fashion by simply doubling their number. External crates and/or their modules can be connected during DSS operation.

#### 4.1.2. The Detector Safety Units

The Detector Safety Unit (DSU) is an assembly of the above mentioned crates installed in a standard 52 U high rack. This assembly contains the infrastructure to connect sensors and actuators via patch panels, as well as display LEDs for monitoring purposes, and also allows the connection to the Ethernet. Each DSU is capable of monitoring is own state by the use of internal sensors and a dedicated I/O module.

---

[6] *O*LE for *P*rocess *C*ontrol
[7] Siemens AG, Germany; www.ad.siemens.com

[8] The DSS uses the CPU 414-4H, the CP443-1 Ethernet communications processor, and the ET 200M external crates with IM153-2 communication modules.
[9] *N*etwork *T*ime *P*rotocol

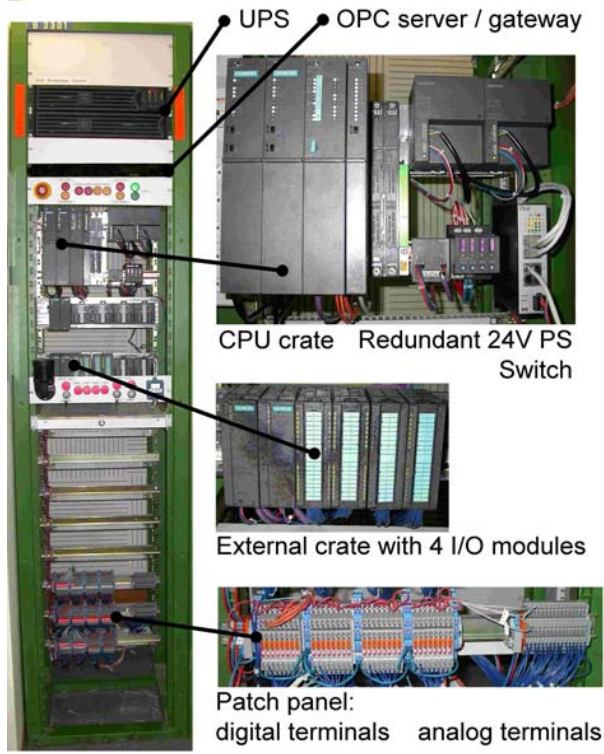

Figure 3: Hardware Layout of a Detector Safety Unit.

At least one DSU is located on the surface and another in the underground experimental cavern, so that the two CPUs can be separated to minimize the danger of accidental damage. The location and the number of DSUs is chosen to minimize the cable installation.
The layout of all DSUs is the same (see

Figure 3). The lower half of the rack is dedicated to the patch panel assembly. The upper half hosts an uninterruptible power supply, the OPC server, the CPU crate and two external crates at maximum (this limits the number of DSUs to 16). Only two DSUs contain a CPU crate, and only one the OPC server.

### 4.1.3. Patch Panels

As a large number of sensors can be connected to the DSU, adequate space for cabling is essential. Twenty-six units of rack space inside of the DSU for the patch panel was found to be the optimal solution for serving two external crates or fifteen I/O modules that can be inserted in one DSU (two external crates times eight slots, minus one slot reserved for the dedicated monitoring module). Each digital module can handle up to 32 channels; each analog module up to eight channels.

The patch panel terminals from WAGO[10] allow for an easy connection of different types of digital and analog sensors. All digital channels are galvanically isolated from each other and from the I/O modules by the use of opto-couplers. Modifications (i.e. adding or removing extra sensors or actuators) can be made during the running of the DSS, and do not interrupt the operation itself.

Threehundredfiftytwo digital sensors and actuators or 120 analog sensors can be connected to one DSU in total. In the case of the digital channels, the limitations are the space constraints inside the rack and the width of the chosen type of WAGO patch panel terminals. Eleven digital I/O modules with up to 32 channels can be handled. The number of analog channels is limited by the number of free slots in the external crates.

All digital sensors and actuators follow the "Positive Safety" rule. The normal condition is signaled by a "high" level (16-24V), while alarm conditions give "low" levels (0-5V). In case of a broken wire, the PLC automatically assumes an alarm condition. Actuators are powered through normally/closed contact relays, which open in the case of an alarm. Short-circuits or broken wires in the analog sensors are detected by abnormal readings.

### 4.1.4. Uninterruptible Power Supplies (UPS)

Each DSU is backed-up by an uninterruptible power supply from APC[11]. Additional battery-packs allow each DSU to be independent of the general power network for at least one hour. This is agreed to be sufficient to bridge the latency until the start of a diesel backed-up power network.

The UPS provides the redundant 24V power supplies of a CPU crate as well as a second pair of redundant 24V power supplies, which deliver the current to all external crates and to the sensors and actuators. A power distribution module allows for current limiting to different sets of consumers.

## 4.2. The OPC Server and Communication

The communication between Front-end and Back-end is routed through a dedicated 1 U high rack-mounted PC[12] acting as a gateway. In addition to a 3COM network card, this PC possesses a special Siemens Ethernet adapter (CP1613) to handle the redundancy of the CPUs: this adapter communicates with both CPUs using ISO protocol. Thus the twofold communication path from the CPUs to the PC is hidden to any external user.

The communication is performed on a dedicated DSS network (DSS COM) using common Ethernet switches.

---

[10] WAGO Kontakttechnik GmbH, Germany; www.wago.com
[11] American Power Conversion Corp. (APC), U.S.A.; www.apc.com. The DSS uses the APC Smart UPS online 1000VA (APC SUOL1000XLI) with one battery packet (APC SUOL48XLBP) and relay I/O module (APC AP-9610).
[12] Supermicro Computer Inc., U.S.A.; www.supermicro.com. The DSS uses the Supermicro SuperServer 5013G-i.

The communication with the Back-end uses the standard CERN network (CERN LAN) connected to the 3COM card.

The data exchange with the Back-end follows the OPC standard [2]. All information exchanged between the PLC and the Back-end is handled by an OPC server running on the gateway PC.

### 4.3. Back-end Architecture

The Back-end acts as user interface for the operators in the experiment control rooms and will display the information, acquired via the Front-end, as well as supplementary information acquired via the network. In addition, it will also show the procedures to be followed, as well as help screens. Furthermore, it will also assist the experts in modifying the safety-related behavior of the Front-end.

The Back-end is implemented using the PVSS SCADA system. Although all safety-critical aspects are concentrated in the Front-end part, the Back-end configuration functionality also has a role for safety in limiting the possibilities for operator input errors. This is carried out via an expert-system-like approach that allows for context analysis and testing to be performed before the download of the new configuration to the Front-end.

The Back-end uses the PVSS database to store the current configuration of the system, while it is planned to store all the modification history into an Oracle Database. This Oracle database will also be used to log the history of all the values of the input channels defined in the system, together with all the alarms detected and the protective actions consequently taken. All the user actions will also be logged.

The operator will be alerted whenever a new alarm is detected. He will have to acknowledge the alarms, and to reset them, once the abnormal situation which triggered the alarm has gone. Subsequently, the operator will be able to take the required actions (probably via the DCS) to bring the detector back into operation.

To ease the work and the understanding of the operator, help pages are foreseen for every alarm, as well as detailed synoptic displays representing the geographical layout of the experiments, and the localization of the alarms and the sensors which generated them.

Finally, the Back-end will also be capable of issuing warnings, so that the operator can react in advance to developing situations that could eventually evolve into full alarms.

### 5. CONCLUSIONS

It has been shown that the initial requirements for the Detector Safety System for the LHC experiments could be fulfilled with a relatively simple and robust system. The prototype of the Front-end part has been built, and the Back-end software part is in progress.

This system will undergo a review in the very near future and, afterwards, the final series production will start. In total about 18 DSUs will be built and deployed to the experimental areas in the coming two years, starting in mid-summer 2003.

The costs for one complete DSS depend largely on the experiment's needs, but are in the order of 50000 to 70000 Euros (sensors, actuators and their cabling are excluded).

### Acknowledgments

We especially want to thank the DSS Working Group[13] [3] for the excellent collaboration on the definition of the requirements. Furthermore, we would like to thank all the people helping with the finalization of the design and the actual construction of the prototype.

---

**THGT007**